\documentclass[twocolumn,showpacs,preprintnumbers,amsmath,amssymb,showkeys,prb]{revtex4}
\usepackage{graphicx}
\usepackage{dcolumn}
\usepackage{bm}

\begin{document}

\title{Quantum transport equations for low-dimensional multiband electronic systems. I}
 \author{I. Kup\v{c}i\'{c}, Z. Rukelj and S. Bari\v{s}i\'{c}}     
 \address{
   Department of Physics, Faculty of Science, University of Zagreb, 
   P.O. Box 331,  HR-10002 Zagreb,  Croatia}

\begin{abstract} 
A systematic method of calculating  the dynamical conductivity tensor in a general multiband electronic model with strong boson-mediated electron-electron interactions is described.
	The theory is based on the exact semiclassical expression for 
the coupling between valence electrons and electromagnetic fields 
and on the self-consistent Bethe--Salpeter equations for the electron-hole propagators.
	The general diagrammatic perturbation expressions for the intraband and interband single-particle
conductivity are determined.
	The relations between the intraband Bethe--Salpeter equation, the quantum transport equation and 
the ordinary transport equation are briefly discussed within the memory-function approximation.
	The effects of the Lorentz dipole-dipole interactions on the dynamical conductivity 
of low-dimensional $sp_\alpha$ models are described in the same approximation.
	Such formalism proves useful in studies of different (pseudo)gapped 
states of quasi-one-dimensional systems with the metal-to-insulator phase transitions and can be easily extended to underdoped two-dimensional high-$T_c$ superconductors.
\end{abstract}
%
%
\maketitle

\section{Introduction}
The ab initio band structure calculations of strongly-interacting low-dimensional (quasi-one-dimensional (Q1D) or two-dimensional (2D)) systems (high-$T_c$ superconductor YBa$_2$Cu$_3$O$_{7-x}$ as well as charge-density-wave (CDW) systems K$_{0.3}$MoO$_3$ and BaVS$_3$ being examples
\cite{Massidda87,Whangbo86,Mattheiss95})
usually reveal several bands in the vicinity of the Fermi level that are, in the leading approximation, decoupled from the rest of the band structure.
	In the tight-binding approximation, the bare dispersions of such a multiband model are well 
understood in terms of the strong hybridization among several properly chosen orbitals per unit cell, while the hybridization with the other electronic levels is neglected.
\cite{Emery87}
	The bare single-particle tight-binding Hamiltonian is thus taken to include 
commensurate/incommensurate CDW or SDW  (spin-density wave) average fields and several first-neighbour bond energies.
	In the two-particle Hamiltonian one usually retains the long-range Coulomb interactions, 
the largest local (Hubbard) interactions together with the most important short-range interactions.
	Associated with different terms in the single-particle Hamiltonian are fluctuations in 
the charge and spin densities on different orbitals in the unit cell and on the corresponding bonds.
	All these ingredients make strongly-interacting multiband electronic problems 
extremely complicated.

To explain most of experimental results on typical single-band systems, it usually suffices to
describe properly the coupling between ${\bf q} \approx 0$ monopole charge fluctuations and external electromagnetic (EM) fields and combine this coupling Hamiltonian with the ordinary transport equations.
\cite{Pines89,Ziman72}
	However, in multiband electronic systems, even in the cases when only one band
intersects the Fermi level, the analysis is considerably more complicated.
	Not only the interband physics but also the intraband physics depends drastically on 
both single-particle and two-particle parameters of the Hamiltonian and are affected differently by various types of fluctuations.
	Therefore, to understand the results of transport measurements
\cite{Ando04,Forro00}
and reflectivity measurements
\cite{Uchida91,Degiorgi91,Kezsmarki06}
of multiband electronic systems in detail, we describe the ${\bf q} \approx 0$ multipolar fluctuations in the dynamical conductivity tensor in a way consistent with the results of experimental methods which probe charge and spin fluctuations which presumably occur at finite wave vectors 
\cite{Sugai03,Takigawa91}.
	The ${\bf q} \approx 0$ charge-charge correlation function is thus just one of several 
equally important correlation functions defined in terms of symmetrized intraband and interband charge and spin fluctuations.
	Common to all these correlation functions is a complicated structure of the intraband 
and interband electron-hole propagators.
	Therefore, to understand the contribution of different types of fluctuations to 
these correlation functions, in particular to the dynamical conductivity, we must first find the solution to quite general Bethe--Salpeter equations for the electron-hole propagators and then combine the result with the very definition of the response function under consideration.
	This question, which is of great importance in studies of strongly-interacting low-dimensional
systems, is in the focus of the present analysis.

In this article, we thus present a general response theory which uses both an essentially exact semiclassical description of the coupling between valence electrons and external EM fields 
\cite{Landau95,Kupcic09b}
and a quite general description of the intraband and interband charge fluctuations
\cite{Barisic90,Kupcic07}.
	The result is the current-dipole Kubo formula for the dynamical conductivity, 
\cite{Kubo95}
with the structure of the intraband electron-hole propagators determined by means of the quantum transport equation (i.e., the intraband Bethe--Salpeter equation).
	To emphasize the connection between the present theory and the ordinary transport theory, 
\cite{Pines89,Ziman72,Eliashberg62,Abrikosov88,Vollhardt80}
we determine the explicit form of the intraband conductivity in the memory-function approximation.
\cite{Gotze72,Giamarchi91,Kupcic04}
	We also find the structure of the interband conductivity in both cases, with real and 
with purely imaginary interband dipole vertices, to show in which way the Lorentz local field corrections
\cite{Adler62,Wieser62,Zupanovic96}
affect the dynamical conductivity of interacting low-dimensional systems.

The article is organized as follows.
	To make the reading of the article easier, we give in section~2 an overview of the results.
	Then in section~3  the notation is introduced to describe electrons and boson modes in a 
general multiband electronic model with boson-mediated electron-electron interactions.
	A complementary multiband electronic model with nonretarded electron-electron interactions
will be studied in a separate article \cite{KupcicUP} which is focused on the commensurability effects in the single-particle conductivity tensor.

Starting with the quantum electrodynamical description of the interaction between valence
electrons and external EM fields, we derive the exact semiclassical form of 
the electron-EM field coupling Hamiltonian in section~4.
	In section~5 we define the dynamical conductivity tensor and briefly discuss the relaxation-time
approximation.
	Section~6 gives the structure of the self-consistent Bethe--Salpeter equations for the dipole
vertex functions and for the related auxiliary electron-hole propagators (to be defined latter) in a general multiband case.
	We show the equivalence of the Bethe--Salpeter equation for the auxiliary intraband
electron-hole propagators and the quantum transport equation.
	The generalized Drude formula is derived from the quantum transport equation
by using the memory-function approximation.
	The bare interband conductivity is considered in section~7 and the related local field 
effects in section~8.
	Section 9 contains the concluding remarks.

\section{Overview of results}
Before discussing the details of the dynamical conductivity model proposed in this article, 
we present an overview of the results.
	In this way principal physical messages of the present work are separated from 
the technical details.

We are interested here in a general low-dimensional model with multiple electronic bands in the presence of boson-mediated electron-electron interactions.
	The single-particle part of the total Hamiltonian is assumed to be exactly solvable.
	The Bloch energies $\varepsilon_L ({\bf k})$ in the band labeled by the band index $L$ 
and all relevant bare vertex functions (the dipole vertices $P_\alpha^{LL'} ({\bf k},{\bf k}')$, for example)
are thus taken as known functions.
	To study the influence of the boson-mediated electron-electron interactions on
single-electron propagators ${\cal G}_L ({\bf k}, {\rm i} \omega_n)$ and, in addition, on different electron-hole propagators encountered in the response functions of interest, we use 
high-order diagrammatic perturbation theory of the metallic state,
as usual in investigations of low-dimensional electronic systems. \cite{Abrikosov75,Dzyaloshinskii73}

The electron-hole propagators are associated with the dispersive, undamped bosons which can either correspond to the external degrees of freedom coupled to electrons/holes or the internal degrees of freedom built of the electron-hole excitations at finite values of {\bf q}.
	In sections~6--8, the emphasis is on the acoustic and Raman-active and infrared-active 
optical phonons.
	The scattering by phase and amplitude phonon modes of the common Peierls CDW model will be studied 
in detail within the microscopic Lee--Rice--Anderson model 
in the accompanying article \cite{KupcicJPCM} (to be referred to as Article II).

  \begin{figure}
\centerline{\includegraphics[width=18pc]{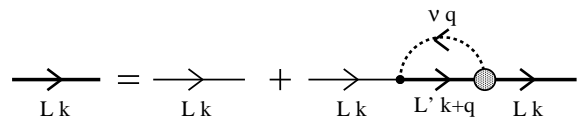}}
   \caption{The Dyson equation (\ref{eq1}) for electron Green's functions.
   $L$ is the electron band index and $\nu$ is the boson branch index.
   The black circle and the big gray circle represent, respectively, the bare and the renormalized
   electron-boson vertex.  
   Here and herafter, the tiny and bold solid (dashed) lines are the bare and renormalized electron 
   (boson) propagators.}
  \end{figure}

In high-order perturbation theory, the electron Green's function ${\cal G}_L ({\bf k}, {\rm i} \omega_n)$ satisfies the Dyson equation
\cite{Fetter71,Abrikosov75,Mahan90}
\begin{eqnarray}
\big[{\rm i} \hbar \omega_n - \varepsilon_L ({\bf k})+\mu - \hbar \Sigma_L ({\bf k}, {\rm i} \omega_n) \big] 
{\cal G}_L ({\bf k}, {\rm i} \omega_n) = \hbar,
\label{eq1}
\end{eqnarray}
where $\Sigma_L ({\bf k}, {\rm i} \omega_n)$ is the self-energy function of the electron
(see figure~1).
	In this case $\Sigma_L ({\bf k}, {\rm i} \omega_n)$ is a function of 
${\cal G}_L ({\bf k}, {\rm i} \omega_n)$, resulting in a system of two coupled equations.
	A strong reduction of the quasi-particle pole weight in ${\cal G}_L ({\bf k}, {\rm i} \omega_n)$ at 
the Fermi level is predicted by different strictly 1D theories
\cite{Dzyaloshinskii73,Solyom79,Giamarchi04}
and observed in angle-resolved photoemission spectroscopy (ARPES) experiments on Q1D systems \cite{Vescoli00,Mitrovic07}.
	In this article we present a systematic method of calculating
the dynamical conductivity tensor in interacting low-dimensional systems which is consistent with 
the described treatment of the single-electron Green's functions ${\cal G}_L ({\bf k}, {\rm i} \omega_n)$.

\subsection{Electron coupling to EM fields}
In semiclassical electrodynamics applied to a general exactly solvable single-particle problem with multiple electronic bands, the electric component of the EM field 
parallel to the cartesian axis $\alpha$, $E_{\alpha} ({\bf q})$, couples to the currents of free charges and to the dipole moments of bound charges in a way described by the coupling Hamiltonian
\begin{equation}
H^{\rm ext} = - \sum_{{\bf q} \alpha} E_{\alpha} ({\bf q}) \hat P_\alpha  ({-\bf q}).
\label{eq2}
 \end{equation} 
	Here, $\hat P_\alpha  ({-\bf q})$ is the total dipole density operator.
	It is the sum of the intraband and interband contributions
\begin{equation}
\hat P_\alpha  ({-\bf q}) =  \sum_{LL'} \sum_{{\bf k} \sigma}
P^{L'L}_{\alpha} ({\bf k}_+,{\bf k})  c_{L'{\bf k}+{\bf q} \sigma}^{\dagger} c_{L{\bf k} \sigma},
\label{eq3}
\end{equation} 
with $P^{L'L}_{\alpha} ({\bf k}_+,{\bf k})$ representing the corresponding intraband ($L=L'$) and interband ($L \neq L'$) dipole vertices, and ${\bf k}_+ = {\bf k}+{\bf q}$.
	It is not hard to verify that these vertices are directly related to the vertices in 
the current and charge density operators, $J_\alpha^{L'L} ({\bf k}_+,{\bf k})$ and 
$q^{L'L} ({\bf k}_+, {\bf k})$.
	The relation is given by the expression
\begin{eqnarray}
P_\alpha^{LL'} ({\bf k},{\bf k}_+) 
= \frac{ {\rm i} \hbar J_\alpha^{LL'} ({\bf k})}{\varepsilon_{L'} ({\bf k}_+) - \varepsilon_L ({\bf k})}
= \frac{{\rm i} e}{q_\alpha}  \, q^{LL'} ({\bf k}, {\bf k}_+)
\label{eq4}
\end{eqnarray}
(${\bf q} = q_\alpha \hat e_\alpha$ in the longitudinal case).
	The imaginary currents and the imaginary dipole moments do not have a straightforward physical 
meaning, but they make the notation used throughout this article as simple as possible.
	For example, they can be used to express the gauge invariance of the expression (\ref{eq2})
simply by rewritting $H^{\rm ext}$ in terms of $q^{L'L} ({\bf k}_+, {\bf k})$ and
$J_\alpha^{L'L} ({\bf k}_+,{\bf k})$ and using the standard relations between 
$E_{\alpha} ({\bf q},\omega)$ and the scalar potential $V^{\rm tot} ({\bf q},\omega)$ and the vector potential $A_{\alpha} ({\bf q},\omega)$.

\subsection{Dynamical conductivity tensor}
In order to examine how electrons respond to applied EM fields, we introduce the dynamical conductivity tensor $\sigma_{\alpha \beta} ({\bf q}, \omega)$, 
\cite{Landau95}
which is the response function relating the induced total current $J_\alpha ({\bf q}, \omega)$ to $E_\beta ({\bf q}, \omega)$;
$J_\alpha ({\bf q}, \omega) = \sum_\beta \sigma_{\alpha \beta} ({\bf q}, \omega)
E_\beta ({\bf q}, \omega)$.
	We use here the finite temperature formalism.
	The dynamical conductivity $\sigma_{\alpha \beta} ({\bf q}, \omega)$ is in this case given by 
analytical continuation (${\rm i} \nu_n \rightarrow \omega + {\rm i} \eta$)
of $\sigma_{\alpha \beta} ({\bf q}, {\rm i} \nu_n)$, which is the Matsubara Fourier transform of 
$\sigma_{\alpha \beta} ({\bf q}, \tau)$.
	The compact form of the semiclassical coupling Hamiltonian (\ref{eq2}) allows us
to express $\sigma_{\alpha \beta} ({\bf q}, \tau)$ in the form of the current-dipole Kubo formula
\cite{Kubo95}
\begin{eqnarray}
\sigma_{\alpha \beta} ({\bf q}, \tau) =  \frac{1}{\hbar V}
\langle  T_\tau [\hat J_\alpha ({\bf q}, \tau) \hat P_\beta (-{\bf q}, 0) ] \rangle_{\rm irred}.
\label{eq5}
\end{eqnarray}
	Here, $\hat J_\alpha ({\bf q})$ is the total current density operator defined 
in a way analogous to (\ref{eq3}) (see (\ref{eq23})).
	In order to extract the most important physical consequences that follow from this
expression, one usually uses the relations (\ref{eq4}) to rewrite 
$\sigma_{\alpha \beta} ({\bf q}, \tau)$ either in terms of the charge-charge or the current-current Kubo formulae.
\cite{Mahan90}
	In contrast to these approaches, we use here (\ref{eq5}), because this 
expression will be shown below to provide the direct link of the present analysis to the common transport theory \cite{Pines89,Ziman72} and to the textbook approaches to the dynamical long-range Coulomb screening \cite{Fetter71,Mahan90,Ziman79,Wooten72}.

   \begin{figure}[tb]
   \centerline{\includegraphics[width=12pc]{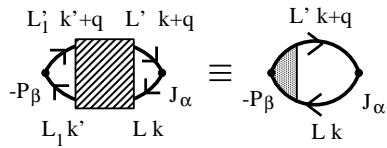}}
   \caption{The Bethe--Salpeter expression for the screened dynamical conductivity tensor 
   $\widetilde \sigma_{\alpha \beta} ({\bf q}, {\rm i} \nu_n)$ in a general multiband case.
   The dynamical conductivity $\sigma_{\alpha \beta} ({\bf q}, {\rm i} \nu_n)$, equation (\ref{eq5}), is the
   RPA-irreducible part of $\widetilde \sigma_{\alpha \beta} ({\bf q}, {\rm i} \nu_n)$. 
   \cite{Kubo95}
   }
   \end{figure}

To complete the response theory, we must take into account the relaxation processes describing the way in which electrons are scattered by static disorder, by other electons or by different types of boson modes.
	Since our primarily interest is in the low-dimensional systems where the quasi-particle
pole weight in the single-electron Green's function is strongly reduced by boson-mediated electron-electron interactions, we must calculate $\sigma_{\alpha \beta} ({\bf q}, \tau)$ beyond the common weak coupling theory (corresponding to
second-order perturbation theory for the single-electron self-energy function).
	This means that in order to work out $\sigma_{\alpha \beta} ({\bf q}, \tau)$ we must use 
the general Bethe--Salpeter expression for the screened conductivity tensor illustrated in figure~2 
and the definition 
$\sum_\beta \sigma_{\alpha \beta} ({\bf q}, \omega) E_\beta ({\bf q}, \omega) = 
\sum_\beta \widetilde \sigma_{\alpha \beta} ({\bf q}, \omega) E_{0\beta} ({\bf q}, \omega)$, where 
${\bf E}_0 ({\bf q}, \omega)$ is the external electric field \cite{Kubo95}.
	The bold solid lines in these two diagrams are the single-electron Green's functions 
which satisfy the Dyson equation (\ref{eq1}).
	The intraband and interband contributions to 
$\sigma_{\alpha \beta} ({\bf q}, {\rm i} \nu_n) = \sum_{LL'} \sigma_{\alpha \beta}^{LL'} ({\bf q}, {\rm i} \nu_n) $ 
can be represented by the same expression,
\begin{eqnarray}
&& \hspace{-10mm} 
\sigma_{\alpha \beta}^{LL'} ({\bf q}, {\rm i} \nu_n) = 
\frac{1}{V} \sum_{{\bf k}  \sigma} \frac{1}{\beta \hbar^2} \sum_{{\rm i} \omega_n} J_{\alpha}^{LL'}({\bf k}) {\cal G}_{L} ({\bf k}, {\rm i} \omega_n )
\nonumber \\
&& \hspace{5mm}
\times  
{\cal G}_{L'} ({\bf k}_+, {\rm i} \omega_{n+})  
(-) \Gamma^{LL'}_{\beta}({\bf k}, {\bf k}_+, {\rm i} \omega_n, {\rm i} \omega_{n+}),
\label{eq6}
\end{eqnarray}
equally well for $L=L'$ as for $L \neq L'$.
	Here the $\Gamma^{LL'}_{\beta}({\bf k}, {\bf k}_+, {\rm i} \omega_n, {\rm i} \omega_{n+})$ are 
the renormalized dipole vertex functions which satisfy the self-consistent Bethe--Salpeter equations for the renormalized dipole vertex, and ${\rm i} \omega_{n+} = {\rm i} \omega_{n} + {\rm i} \nu_{n}$.
	In order to emphasize the connection between the present approach and the common transport
theory, as well as to simplify the notation, we introduce the auxiliary electron-hole propagator
$\Phi^{LL'}_\beta ({\bf k},{\bf k}_+,  {\rm i} \omega_n, {\rm i} \omega_{n+})$ by comparing the relation 
\begin{equation}
\sigma_{\alpha \beta}^{LL'} ({\bf q}, {\rm i} \nu_n) 
= \frac{1}{\beta V}  \sum_{i \omega_n {\bf k} \sigma} J_{\alpha}^{LL'}({\bf k}) 
\Phi^{LL'}_\beta ({\bf k},{\bf k}_+,  {\rm i} \omega_n, {\rm i} \omega_{n+})
\label{eq7}
\end{equation}
with (\ref{eq6}).
	Therefore, our task is to solve the self-consistent equations for 
$\Phi^{LL'}_\beta ({\bf k},{\bf k}_+,  {\rm i} \omega_n, {\rm i} \omega_{n+})$ 
rather than that for the renormalized dipole vertices.

We limit our discussion to the low-dimensional systems where the scattering by static disorder and the scattering processes in which electrons change the band can both be neglected.
	These processes have negligible influence on the properties studied here.
	However, the self-consistent equations still have a very complex structure in particular as 
a consequence of the following features: the nesting properties of the Fermi surface,
the resonant nature of the scattering by the (quasi)static CDW or SDW potentials and
the local-field effects in the interband conductivity channel.

\subsection{Results}
The results of the present analysis are the following:
{\it (i)} we show the formal solution to the Bethe--Salpeter equations for 
$\Phi^{LL'}_\alpha ({\bf k},{\bf k}_+,  {\rm i} \omega_n, {\rm i} \omega_{n+})$ and to the (intraband) quantum transport equation, 
{\it (ii)} we compare the intraband conductivity to that obtained by using the memory-function approximation, 
{\it (iii)} we calculate the bare interband conductivity and
{\it (iv)} we rederive the well-known Lorentz--Lorenz form of the optical conductivity in the presence of 
scattering by boson modes.

{\it (i)} Although it is tempting to completely neglect 
the electron-EM field vertex corrections
in $\sigma_{\alpha \beta} ({\bf q}, {\rm i} \nu_n)$, 
this is not justified in the intraband channel, at least in the cases with low-dimensional conductivity studied here.
	The method by which the intraband conductivity tensor is calculated
here is very general and is based on a systematic analysis of 
the self-consistent equation for the auxiliary electron-hole propagator
$\Phi_\beta ({\bf k},{\bf k}_+,  {\rm i} \omega_n, {\rm i} \omega_{n+})$ in which the single-electron self-energy contributions and the related vertex corrections are treated on an equal footing.
	The equation is of the form 
\begin{eqnarray}
&& \hspace{-5mm}
\big[{\rm i} \nu_{n} + \varepsilon({\bf k})/\hbar -\varepsilon({\bf k}_+)/\hbar
\nonumber \\
&&  \hspace{10mm}
+  \Pi ({\bf k},{\bf k}_+, {\rm i} \omega_n,{\rm i} \omega_{n+})  \big]
\Phi_\alpha ({\bf k},{\bf k}_+, {\rm i} \omega_n, {\rm i} \omega_{n+} ) 
\nonumber \\
&&  \hspace{5mm}
= \frac{1}{\hbar^2} \big[ {\cal G} ({\bf k}, {\rm i} \omega_n ) 
- {\cal G} ({\bf k}_+, {\rm i} \omega_{n+} ) \big] p_\alpha (-{\bf q}),
\label{eq8}
\end{eqnarray}
with $p_\alpha (-{\bf q}) = {\rm i} e/q_\alpha$ and
$ \Pi ({\bf k},{\bf k}_+, {\rm i} \omega_n,{\rm i} \omega_{n+}) =  
\Pi^r ({\bf k},{\bf k}_+, {\rm i} \omega_n,{\rm i} \omega_{n+}) + 
{\rm i}  \Pi^i ({\bf k},{\bf k}_+, {\rm i} \omega_n, {\rm i} \omega_{n+})$.
	For simplicity we assume that only one band intersects the Fermi level and 
omit explicit reference to the conduction band index.
	$\Pi ({\bf k},{\bf k}_+, {\rm i} \omega_n,{\rm i} \omega_{n+})$ is the electron-hole self-energy, the
structure of which is derived here for the case in which conduction electrons are scattered only by various boson modes.
	These boson modes are assumed to dissipate momentum by their own means (impurities, Umklapps, etc.).
	Pinned phasons of the commensurate CDW problem, for example, enter in the present theory 
in the same way as the infrared-active optical phonons from section 8.2.
	This question is discussed in more details in Article II.

The electron-hole self-energy is found to be the difference between two auxiliary single-electron
self-energies $\widetilde \Sigma ({\bf k}, {\rm i} \omega_n )$, and 
the contributions of the ${\bf q} \approx 0$ forward scattering processes are found to drop out of
both $\widetilde \Sigma ({\bf k}, {\rm i} \omega_n )$ and $\Pi ({\bf k},{\bf k}_+, {\rm i} \omega_n,{\rm i} \omega_{n+})$, in full agreement with the charge continuity equation.

Equation (\ref{eq8}) is valid under quite general conditions, and is therefore applicable to a rich variety of problems, including purely electronic models with nonretarded local and short-range interactions.
	For example, it is not hard to verify 
that the normal scattering processes  do not lead to the resistivity in this case, as required by the general principle of microscopic reversibility. \cite{Ziman72}
	This question will be discussed in more details in Ref.~\onlinecite{KupcicUP}.

{\it (ii)} 
In the ordinary 3D metallic regime, equation~(\ref{eq8}) leads to the ordinary transport equation, with \cite{Pines89,Ziman72}
\begin{equation}
\sigma^{\rm intra}_{\alpha \alpha} ({\bf q},\omega) \approx \frac{{\rm i} e^2}{m} \frac{1}{V} \sum_{{\bf k} \sigma}
\frac{m v^2_\alpha ({\bf k}) \big(- \partial f({\bf k})/\partial E ({\bf k})\big)}{
\omega + {\rm i} \Gamma_{1\alpha} ({\bf k}) -q_\alpha^2 v^2_\alpha ({\bf k})/\omega}
\label{eq9}
\end{equation}
being the resulting intraband electrical conductivity, and ${\bf q} = q_\alpha \hat e_\alpha$ again.
	Here, $m$ is the unity of mass, commonly taken as the free electron mass,
and $\Gamma_{1\alpha} ({\bf k}) \propto \Pi^i ({\bf k},{\bf k}_+, {\rm i} \omega_n, {\rm i} \omega_{n+})$ 
is the intraband relaxation rate.
	In the Drude $\omega \approx 0$ limit, where $\omega \gg q_\alpha v_\alpha ({\bf k})$, 
the $q_\alpha$ dependent term in the denominator of (\ref{eq9}) can be ignored,
and we obtain the intraband conductivity (\ref{eq10}).
	For $\omega \ll q_\alpha v_\alpha ({\bf k})$ and $q_\alpha \rightarrow 0$, on the other hand, 
the result is the well-known Thomas--Fermi expression  \cite{Ziman79,Mahan90} for 
$\sigma^{\rm intra}_{\alpha \alpha} ({\bf q},\omega)$
and for the dielectric susceptibility $\chi^{\rm intra} ({\bf q},\omega) = (q_\alpha^2 / {\rm i} \omega)
\sigma^{\rm intra}_{\alpha \alpha} ({\bf q},\omega)$.

{\it (iii)} 
The general expression for the interband single-particle conductivity is derived and compared to the approximate expression which is frequently encountered in the literature. \cite{Wooten72,Ziman79}

   \begin{figure}[tb]
   \centerline{\includegraphics[width=19pc]{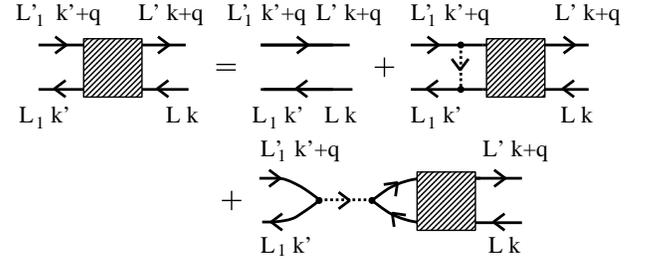}}
   \caption{The Bethe--Salpeter equations for electron-hole propagators 
   in a general multiband electronic model with weak 
   nonretarded and/or boson-mediated electron-electron interactions.
   The vertex renormalizations are assumed to be negligible in this case.
   }
   \end{figure}

{\it (iv)} 
If the scattering of valence electrons is sufficiently weak, the electron-boson vertex renormalizations are of no importance and the electron-hole propagators of figure~2 satisfy the Bethe--Salpeter equations shown in figure~3 with two types of contributions, 
the ladder contributions and the RPA (random phase approximation) contributions.
\cite{Negele88}
	These equations can be completely rewritten in terms of the renormalized single-electron 
Green's functions and the dynamically screened electron-electron interactions.

This approximation is used here to illustrate how the conductivity tensor is affected by
the short-range dipole-dipole interactions.
	The result is the well-known Lorentz-Lorenz form of the dynamical conductivity
for 3D metallic $sp_\alpha$ models \cite{Adler62,Wieser62} which is believed to be valid in the Q1D conductors, as well \cite{Zupanovic96}.
	From the result
\begin{eqnarray}
\sigma^{\rm intra }_{\alpha \alpha} ({\bf q},\omega) = \sigma^{\rm intra, 0 }_{\alpha \alpha} ({\bf q},\omega)
= \frac{{\rm i} e^2 n_{\alpha \alpha}^{\rm eff}/m}{\omega + {\rm i} \Gamma_{1\alpha}},
\label{eq10} \\
\sigma^{\rm inter}_{\alpha \alpha} ({\bf q},\omega) = 
\frac{\sigma^{\rm inter,0}_{\alpha \alpha} ({\bf q},\omega) }{
1 - (4\pi {\rm i}/3 \omega) \sigma^{\rm inter,0}_{\alpha \alpha} ({\bf q},\omega)},
\label{eq11} 
\end{eqnarray}
we can conclude that the contributions to the conductivity tensor of the free charges and of the bound charges are decoupled and that the Lorentz dipole-dipole interactions affect only the bound (interband) charge contribution to the conductivity tensor.

\section{Electron-boson coupling}
Let us first introduce the quantities characterizing a general multiband electronic model with boson-mediated electron-electron interactions.
	The total Hamiltonian is the sum of two contributions, $H = H_0 + H'$.
	Here,
\begin{eqnarray}
&& \hspace{-10mm}
H_0 = H_0^{\rm el} + H_0^{\rm bos} =
\sum_{L {\bf k} \sigma} \varepsilon_L ({\bf k}) c^{\dagger}_{L{\bf k} \sigma} c_{L{\bf k} \sigma}
\nonumber \\
&& \hspace{-2mm}
+  \sum_{\nu {\bf q}'} \frac{1}{2M_\nu} \big[  p^{\dagger}_{\nu {\bf q}'} p_{\nu {\bf q}'} + \big( 
M_\nu \omega_{\nu {\bf q}'} \big)^2 q^{\dagger}_{\nu {\bf q}'} q_{\nu {\bf q}'}\big]
\label{eq12}
 \end{eqnarray} 
is the bare electron-boson Hamiltonian, and 
\begin{equation}
H' = \sum_{\nu {\bf q}'} \sum_{L L' {\bf k} \sigma} \Phi^{L'L}_\nu ({\bf k}+{\bf q}', {\bf k})
c^{\dagger}_{L'{\bf k}+{\bf q}'  \sigma} c_{L{\bf k} \sigma}
\label{eq13}
\end{equation}
is the scattering Hamiltonian which
describes the scattering of valence electrons by each of the boson modes.
	As already mentioned, the bosonic modes can be external to electronic system (e.g. phonons) 
or built from electron-hole fluctuations at finite wave vectors.
	In (\ref{eq12}), $\varepsilon_L ({\bf k})$ is the bare electron dispersion,  
$\omega_{\nu {\bf q}'}$ is the bare boson frequency and $M_\nu$ is the corresponding ``mass'' parameter.
	Finally,  $\Phi^{L'L}_\nu ({\bf k}+{\bf q}', {\bf k})$ is the generalized 
scattering potential.
	For convenience, it is assumed to be a product of the coupling constant $g_\nu/\sqrt{N}$, 
the dimensionless electron-boson vertex $q^{L'L}_\nu ({\bf k}+{\bf q}',{\bf k})$ and the boson field $q_{\nu {\bf q}'}$,
\begin{equation}
\Phi^{L'L}_\nu ({\bf k}+{\bf q}', {\bf k}) = \frac{g_\nu}{\sqrt{N}} \,
 q^{L'L}_\nu ({\bf k}+{\bf q}',{\bf k}) q_{\nu {\bf q}'}.
\label{eq14}
 \end{equation}
	In the intraband scattering approximation, in which the electron does not change the band when 
it is scattered by bosons, the scattering potential $\Phi^{L'L}_\nu ({\bf k}+{\bf q}', {\bf k})$ is diagonal in the band index.

\subsection{Skeleton diagrams}

  \begin{figure}
   \centerline{\includegraphics[width=18pc]{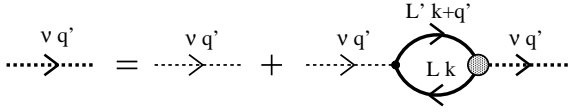}}
   \caption{The Dyson equation (\ref{eq16}) for boson Green's functions.
   }
  \end{figure}

In the finite temperature formalism, the effects of $H'$ on the single-electron and boson properties are described in terms of two exact Matsubara Green's functions
\cite{Fetter71,Abrikosov75,Mahan90}
\begin{eqnarray}
&& \hspace{-10mm}
{\cal G}_L ({\bf k}, \tau) = - \langle T_\tau \big[c_{L{\bf k} \sigma}(\tau) 
c^{\dagger}_{L{\bf k} \sigma}(0) \big] \rangle, 
\nonumber \\
&& \hspace{-11mm}
{\cal D}_\nu ({\bf q}', \tau) = - \langle T_\tau \big[q_{\nu {\bf q}'}(\tau) 
q^\dagger_{\nu {\bf q}'}(0) \big] \rangle.
\label{eq15} 
\end{eqnarray}
	Their Fourier transforms satisfy two Dyson equations, equation~(\ref{eq1}) and 
\begin{equation}
\big\{ M_\nu \big[({\rm i} \nu_{n})^2 - \omega_{\nu{\bf q}'}^2\big] 
- \hbar \Sigma_\nu ({\bf q}', {\rm i} \nu_n) \big\} {\cal D}_\nu ({\bf q}', {\rm i} \nu_{n}) = \hbar
\label{eq16}
\end{equation}
(see figure~4).
	It is convenient to adopt an abbreviated notation by writting
\begin{equation}
\frac{g_\nu^2}{N} |q_\nu^{L'L} ({\bf k}+{\bf q}',{\bf k})|^2  {\cal D}_\nu ({\bf q}',  \tau)
\equiv \hbar {\cal F}_\nu^{L'L} ({\bf k}+{\bf q}',{\bf k},  \tau).
\label{eq17}
\end{equation}
	This expression can also be shown in the form 
${\cal F}_\nu^{L'L} ({\bf k}+{\bf q}',{\bf k}, \tau) = 
{\cal F}_\nu^{L'L,LL'} ({\bf k}+{\bf q}',{\bf k}, {\bf q}',\tau)$, where 	
\begin{eqnarray}
&& \hspace{-10mm}
\hbar {\cal F}_\nu^{L'L,L_1L_1'} ({\bf k}+{\bf q}',{\bf k}', {\bf q}',\tau) 
\nonumber \\ 
&& \hspace{-5mm}
= - \langle T_\tau \big[ \Phi^{L'L}_\nu ({\bf k}+{\bf q}', {\bf k}, \tau)
\Phi^{L_1L_1'}_\nu ({\bf k}',{\bf k}'+{\bf q}', 0) \big] \rangle,
\label{eq18}
\end{eqnarray}
is by definition the force-force correlation function \cite{Mahan90,Kupcic09} (FFCF) associated with the potential $\Phi^{L'L}_\nu ({\bf k}+{\bf q}', {\bf k})$.

  \begin{figure}[tb]
  \centerline{\includegraphics[width=20pc]{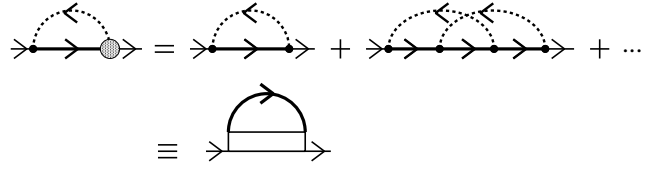}}
  \caption{The skeleton series for the single-electron self-energy.
  The self-energy is shown in terms of both the renormalized three-point 
  vertex function $\tilde q_\nu ^{L'L} ({\bf k}_+,{\bf k}, {\rm i} \omega_{n+}, {\rm i} \omega_{n})$
  (the big gray circle) and the completely irreducible four-point vertex function 
  $U_\nu ^{L'L,LL'} ({\bf k}+{\bf q}', {\bf k},{\bf q}', {\rm i} \omega_{n+}, {\rm i} \omega_{n}, 
  {\rm i} \nu_{n})$ 
  (the white rectangle).
  }
  \end{figure}

In high-order perturbation theory the self-energy function of the electron, 
$\Sigma_L ({\bf k}, {\rm i} \omega_n )$, is usually represented by the sum of skeleton diagrams shown in figure~5, leading to a simple analytical expression 
\begin{eqnarray}
&& \hspace{-10mm}
\hbar \Sigma_L ({\bf k}, {\rm i} \omega_n ) \approx  -\sum_{L' \nu {\bf q}'} \frac{1}{\beta \hbar} 
\sum_{{\rm i}  \nu_n} {\cal G}_{L'} ({\bf k}+{\bf q}', {\rm i} \omega_{n+})
\nonumber \\
&& \hspace{5mm} \times 
\widetilde {\cal F}_\nu^{L'L,LL'} ({\bf k}+{\bf q}',{\bf k}, {\bf q}', {\rm i} \omega_{n+}, {\rm i} \omega_{n}).
\label{eq19}
\end{eqnarray}
	Here, 
$\widetilde {\cal F}_\nu^{L'L,LL'} ({\bf k}+{\bf q}',{\bf k}, {\bf q}', {\rm i} \omega_{n+}, {\rm i} \omega_{n})$ 
is the renormalized FFCF obtained  by replacing one bare electron-boson vertex in 
${\cal F}_\nu^{L'L,LL'} ({\bf k}+{\bf q}',{\bf k}, {\bf q}',{\rm i} \nu_{n})$,
$q_\nu ^{L'L} ({\bf k}_+,{\bf k})$, by the renormalized vertex
$\tilde q_\nu ^{L'L} ({\bf k}_+,{\bf k}, {\rm i} \omega_{n+},{\rm i} \omega_{n})$.
	Equation (\ref{eq19}) is nothing but the three-point vertex representation of 
$\Sigma_L ({\bf k}, {\rm i} \omega_n )$.
	In the same representation the boson self-energy takes the form
\begin{eqnarray}
&& \hspace{-10mm}
\hbar \Sigma_\nu ({\bf q}, {\rm i} \nu_n ) \approx \sum_{LL'{\bf k} \sigma } \frac{1}{\beta \hbar^2} \sum_{{\rm i}  \omega_n}
\frac{g_\nu^2}{N}  \tilde q_\nu ^{L'L} ({\bf k}_+,{\bf k}, {\rm i} \omega_{n+},{\rm i} \omega_{n})  
\nonumber \\
&& \hspace{10mm} \times 
q_\nu ^{LL'} ({\bf k},{\bf k}_+) {\cal G}_{L} ({\bf k}, {\rm i} \omega_{n}) 
{\cal G}_{L'} ({\bf k}_+, {\rm i} \omega_{n+}).
\label{eq20}
\end{eqnarray}

We have just reemphasized that the three-point vertex representation is very useful in describing the closed system of four equations characterizing strongly-interacting
electron-boson systems in the absence of external EM fields (equations~(\ref{eq1}), (\ref{eq16}), (\ref{eq19}) and (\ref{eq20})).
	However, the conductivity tensor of such an electron-boson model has a compact form if the 
structure of the Bethe--Salpeter equations of section 6 is also simple, at least from the formal point of view.
	For this purpose, it is much more convenient to use the four-point vertex representation.
	This representation is based upon the knowledge of
the irreducible four-point interaction 
$U_\nu ^{L'L,LL'} ({\bf k}+{\bf q}', {\bf k},{\bf q}', {\rm i} \omega_{n+}, {\rm i} \omega_{n}, 
{\rm i} \nu_{n})$.
	The equivalence of the two representations in the skeleton approximation is illustrated
in figure~5, where 
$U_\nu ^{L'L,LL'} ({\bf k}+{\bf q}', {\bf k},{\bf q}', {\rm i} \omega_{n+}, {\rm i} \omega_{n}, 
{\rm i} \nu_{n})$
is represented by the white rectangle.
	The expression for $\Sigma_L ({\bf k}, {\rm i} \omega_n )$ is now of the form
\begin{eqnarray}
&& \hspace{-12mm}
\hbar \Sigma_L ({\bf k}, {\rm i} \omega_n ) \approx  -\sum_{L' \nu {\bf q}'} \frac{1}{\beta \hbar} 
\sum_{{\rm i}  \nu_n} {\cal G}_{L'} ({\bf k}+{\bf q}', {\rm i} \omega_{n+})
\nonumber \\
&& \hspace{0mm} \times 
U_\nu ^{L'L,LL'} ({\bf k}+{\bf q}', {\bf k},{\bf q}', {\rm i} \omega_{n+}, {\rm i} \omega_{n}, 
{\rm i} \nu_{n}).
\label{eq21}
\end{eqnarray}
	A full discussion of the Bethe--Salpeter equations is postponed to section 6.

\section{Electron coupling to EM fields}
In quantum electrodynamics of tightly bound valence electrons in solids, whether metallic or insulating,
the coupling between electrons and external EM fields is given by the minimal gauge invariant substitution.
	The expansion to the second order in the vector potential $A_\alpha ({\bf q}, \omega)$ gives
the coupling Hamiltonian
$H^{\rm ext} \approx \widetilde H^{\rm ext}_1 + \widetilde H^{\rm ext}_2$, with \cite{Kupcic07,Kupcic04,Kupcic05}
\begin{eqnarray}
 && \hspace{-10mm}
\widetilde H^{\rm ext}_1  =  -\frac{1}{c} \sum_{{\bf q} \alpha} A_{\alpha} ({\bf q})
\hat J_{\alpha} (-{\bf q}),
\nonumber \\
&& \hspace{-10mm}
\widetilde H^{\rm ext}_2 =  \frac{e^2}{2mc^2} \sum_{{\bf q} {\bf q}'\alpha \beta} 
A_{\alpha} ({\bf q}-{\bf q}')A_{\beta} ({\bf q}') \hat  \gamma^{LL}_{\alpha \beta}(-{\bf q};2).
\label{eq22}
 \end{eqnarray} 
	Here,
\begin{eqnarray}
&& \hspace{-10mm}
\hat J_\alpha  ({\bf q}) = \sum_{LL'}\sum_{{\bf k} \sigma} J_\alpha^{LL'} ({\bf k}, {\bf k}_+)
c^\dagger_{L{\bf k} \sigma} c_{L'{\bf k}+{\bf q} \sigma},
\nonumber \\
&& \hspace{-15mm}
\hat  \gamma^{LL}_{\alpha \beta}({\bf q};2)= \sum_{L}\sum_{{\bf k} \sigma} 
\gamma^{LL}_{\alpha \beta}({\bf k}, {\bf k}_+;2) c^\dagger_{L{\bf k} \sigma} c_{L{\bf k}+{\bf q} \sigma}
\label{eq23}
\end{eqnarray}
are, respectively, the total current density operator and  the bare diamagnetic density operator.

To obtain the semiclassical expression for the coupling Hamiltonian, equation~(\ref{eq2}), we have 
to take into account the fact that both the interband and intraband current-current contributions to the diamagnetic current must vanish in the normal metallic or insulating state.
	The cancellation of the interband contribution is a consequence of the effective mass theorem,
\cite{Abrikosov73,Kupcic07,Kupcic04,Kupcic03}
\begin{eqnarray}
&& \hspace{-10mm}
\gamma^{LL}_{\alpha \beta}({\bf k}) = \gamma^{LL}_{\alpha \beta}({\bf k};2)
+ \sum_{L'(\neq L)} \frac{(2 m/e^2) J_\alpha^{L L'}({\bf k}) J_\beta^{L'L}({\bf k})
}{\varepsilon_L({\bf k}) - \varepsilon_{L'}({\bf k})}
\nonumber \\
&& \hspace{2mm}
= \frac{m}{\hbar^2} \frac{\partial^2  \varepsilon_L ({\bf k})}{\partial k_\alpha \partial k_\beta},
\label{eq24}
 \end{eqnarray} 
leading to $H^{\rm ext} \approx  H^{\rm ext}_1 +  H^{\rm ext}_2$, with \cite{Kupcic09}
\begin{eqnarray}
&& \hspace{-5mm}
H^{\rm ext}_1 = -\frac{1}{c} \sum_{{\bf q} \alpha} A_{\alpha} ({\bf q}) 
\bigg[{\sum_L}\hat{J}^{LL}_{\alpha} (-{\bf q})
+\sum_{L\neq L'} {\rm i} \omega\hat{P}^{LL'} _{\alpha} (-{\bf q}) \bigg], 
\nonumber \\
&& \hspace{-5mm}
H^{\rm ext}_2 =  \frac{e^2}{2mc^2} \sum_{L {\bf q} {\bf q}'\alpha \beta} 
A_{\alpha} ({\bf q}-{\bf q}')A_{\beta} ({\bf q}') 
\hat{\gamma}^{LL}_{\alpha \beta} (-{\bf q}). 
\label{eq25}
 \end{eqnarray} 
	Here, $\hat P_\alpha^{LL'}  ({\bf q})$, $L \neq L'$, is the interband part in dipole 
density operator (\ref{eq3}), and $\hat{\gamma}^{LL}_{\alpha \beta} (-{\bf q})$ is the diamagnetic 
density operator.
	The reciprocal effective mass tensor $\gamma^{LL}_{\alpha \beta}({\bf k})$ is 
the vertex function in  $\hat{\gamma}^{LL}_{\alpha \beta} (-{\bf q})$.

The intraband current-current contribution to the diamagnetic current cancels $H^{\rm ext}_2$, 
confirming the assertion made in section 2 that
\begin{eqnarray}
H^{\rm ext} = - \sum_{{\bf q} \alpha} E_{\alpha} ({\bf q}) \hat P_\alpha  ({-\bf q}).
\nonumber
 \end{eqnarray} 
	We emphasize the simple structure of this coupling Hamiltonian.
	The most direct way to obtain this expression for $H^{\rm ext}$ is to consider 
the coupling of the scalar potential $V^{\rm tot} ({\bf q}, \omega)$ to the charge density operator. \cite{Kubo95}

\subsection{Density operators}
Let us derive now the relations (\ref{eq4}) among three types of intraband and interband vertex functions underlying the gauge invariance of the present response theory.
	To do this, we use very general microscopic operator equations
\begin{eqnarray}
&& \hspace{-10mm}
 \hbar q_\alpha \hat J_\alpha({\bf q}) = {\rm i} \hbar \frac{\partial}{\partial t} 
\hat \rho ({\bf q}) = \big[\hat \rho({\bf q}), H_0^{\rm el} \big],
\nonumber \\
&& \hspace{-10mm}
{\rm i} \hbar \hat J_\alpha({\bf q}) = {\rm i} \hbar \frac{\partial}{\partial t} 
\hat P_\alpha({\bf q}) = \big[\hat P_\alpha({\bf q}), H_0^{\rm el} \big],
\label{eq26}
\end{eqnarray}
where $H_0^{\rm el}$ is the electronic part in the bare Hamiltonian (\ref{eq12}).
	Furthermore,
\begin{eqnarray}
\hat \rho ({\bf q}) = \sum_{LL'}\sum_{{\bf k} \sigma} e q^{LL'} ({\bf k}, {\bf k}_+)
c^\dagger_{L{\bf k} \sigma} c_{L'{\bf k}+{\bf q} \sigma}
\label{eq27}
\end{eqnarray}
is the total charge density operator,
with $e q^{LL'} ({\bf k}, {\bf k}_+)$ being the charge vertex function, while 
$\hat J_\alpha ({\bf q})$ and $\hat P_\alpha ({\bf q})$ are given by (\ref{eq23}) and (\ref{eq3}).
	The two commutators in (\ref{eq26})  are easily worked out, leading to the 
required relations
\begin{eqnarray}
P_\alpha^{LL'} ({\bf k},{\bf k}_+) 
= \frac{ {\rm i} \hbar J_\alpha^{LL'} ({\bf k})}{\varepsilon_{L'} ({\bf k}_+) - \varepsilon_L ({\bf k})}
= \frac{{\rm i} e}{q_\alpha}  \, q^{LL'} ({\bf k}, {\bf k}_+).
\nonumber
\end{eqnarray}
	The explicit form of each of these vertices is often obtained by calculating  the coefficients
in the linear term in $H^{\rm ext}$ in (\ref{eq22}).

The long-wavelength intraband vertices are of particular interest.
	They are given by model independent expressions
\begin{eqnarray}
&& \hspace{-10mm}
q^{LL} ({\bf k}, {\bf k}_+) \approx 1, 
\nonumber \\
&& \hspace{-10mm}
J^{LL}_\alpha ({\bf k}, {\bf k}_+) \approx J^{LL}_\alpha ({\bf k}) \equiv e v^{L,0}_\alpha ({\bf k}),
\nonumber \\
&& \hspace{-10mm}
P^{LL}_\alpha ({\bf k}, {\bf k}_+) \approx P^{LL}_\alpha (-{\bf q}) \equiv p_\alpha (-{\bf q}).
\label{eq28}
\end{eqnarray}
	Here, 
$v^{L,0}_\alpha ({\bf k}) =  (1/\hbar) \partial \varepsilon_L ({\bf k})/\partial  k_\alpha$ 
is the bare electron group velocity and $p_\alpha (-{\bf q}) = {\rm i} e/q_\alpha$ is the bare dipole vertex.

Notice that the thermodynamic averages of the density operators in (\ref{eq26}) are the usual induced macroscopic densities.
	For example, ${\bf J} ({\bf q}, \omega)$, 
$\rho ({\bf q}, \omega)$ and ${\bf P} ({\bf q}, \omega)$ represent the common notation for the induced macroscopic current, charge and dipole densities.
	They are related by two equations
\cite{Landau95}
\begin{eqnarray}
&&\hspace{-15mm}
{\bf q} \cdot {\bf J} ({\bf q}, \omega) = \omega \rho ({\bf q}, \omega),
\nonumber \\
&&\hspace{-10mm}
\rho ({\bf q}, \omega) = -{\rm i} {\bf q} \cdot {\bf P} ({\bf q}, \omega).
\label{eq29}
\end{eqnarray}
	The first one is the usual charge continuity equation.

\section{Dynamical conductivity tensor}
We now define the total dynamical conductivity tensor $\sigma_{\alpha \beta} ({\bf q}, \omega)$
as the response function to the macroscopic electric field 
\cite{Landau95}
\begin{eqnarray}
{\bf E} ({\bf q}, \omega) = {\bf E}_0 ({\bf q}, \omega) + {\bf E}_1 ({\bf q}, \omega);
\label{eq30}
\end{eqnarray}
that is
\begin{eqnarray}
J_{\alpha} ({\bf q}, \omega) = \sum_\beta \sigma_{\alpha \beta} ({\bf q}, \omega) 
E_{\beta} ({\bf q}, \omega).
 \label{eq31}
\end{eqnarray}
	Here, ${\bf J} ({\bf q}, \omega)$ is the induced total current density from (\ref{eq29}),
${\bf E}_0 ({\bf q}, \omega)$ is the external field and ${\bf E}_1 ({\bf q}, \omega)$ is the depolarization field (equal to $-4\pi {\bf P} ({\bf q}, \omega)$ for the longitudinal polarization of the field in the thin slab).
	The conductivity tensor is given by analytical continuation to the real axis of  
$\sigma_{\alpha \beta} ({\bf q}, {\rm i} \nu_n)$ (see figure~6), where 
$\sigma_{\alpha \beta} ({\bf q}, {\rm i} \nu_n)$ is the Matsubara Fourier transform of 
\begin{eqnarray}
\sigma_{\alpha \beta} ({\bf q}, \tau) =  \frac{1}{\hbar V}
\langle  T_\tau [\hat J_\alpha ({\bf q}, \tau) \hat P_\beta (-{\bf q}, 0) ] \rangle_{\rm irred}.
\nonumber  
\end{eqnarray}
	Therefore, to find the dynamical conductivity tensor of an interacting electron-boson 
system, we must determine the structure of the current-dipole Kubo formula (\ref{eq5}) which  agrees with the definition of the macroscopic electric field (\ref{eq30}).

\begin{figure}[tb]
\centerline{\includegraphics[width=17pc]{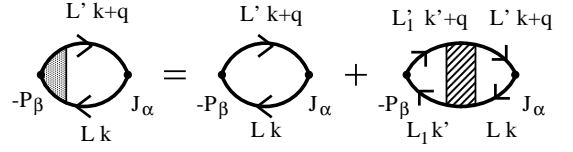}}
\caption{
   The dynamical conductivity $\sigma_{\alpha \beta} ({\bf q}, {\rm i} \nu_n)$ from figure~2, 
   shown in the three-point vertex representation 
   (left diagram) and in the four-point vertex  representation (right diagrams).
   The collection of diagrams in the renormalized four-point interaction must be consistent with 
   the definition of ${\bf E} ({\bf q}, \omega)$ in (\ref{eq30}); 
   i.e., the renormalized four-point
   interaction is irreducible in the long-range RPA contribution.
}
\end{figure}

\subsection{Relaxation-time approximation}
It is instructive first to show the structure of the bare dynamical conductivity (\ref{eq5}) in the simplest multiband model, namely, a two-band model with the lower band partially occupied by electrons and the upper band empty, in the limiting case where the scattering is sufficiently weak.
	As explained in more details in Article II, 
this model is of importance in accounting for the properties of the ordered state of some simple CDW systems.
\cite{Lee74,Kim93,Kupcic09}
	
To do this, we first calculate the ideal bare conductivity tensor 
$\sigma^0_{\alpha \beta} ({\bf q}, {\rm i} \nu_n)$ which is given by the first term in figure~6, with 
${\cal G}_L({\bf q}, {\rm i} \omega_n)$ replaced by the bare Green's function 
${\cal G}_L^0({\bf q}, {\rm i} \omega_n)$.
	Summation over ${\rm i} \omega_n$ can be performed easily to give 
$\sigma^0_{\alpha \beta} ({\bf q}, {\rm i} \nu_n) = \delta_{\alpha, \beta} \sum_{LL'}
\sigma_{\alpha \alpha}^{LL'} ({\bf q}, {\rm i} \nu_n)$.
	After analytical continuation, we obtain the ideal bare intraband ($L=L'$) and interband
($L\neq L'$) contributions described by 
\begin{eqnarray}
&& \hspace{-12mm}  
\sigma_{\alpha \alpha}^{LL'} ({\bf q}, \omega) 
 \nonumber \\
 && \hspace{-7mm} 
=  -\frac{1}{V}  \sum_{{\bf k} \sigma} 
\frac{J_{\alpha}^{LL'}({\bf k}) P_{\alpha}^{L'L}({\bf k}) [f_{L}({\bf k})- f_{L'}({\bf k}_+)]}{
\hbar \omega + {\rm i} \hbar \eta  + \varepsilon_{L}({\bf k})- \varepsilon_{L'}({\bf k}_+) }.
\label{eq32}
\end{eqnarray}
	Weak scattering processes can be inserted into this expression
by replacing the adiabatic term $\eta$ by the relaxation rate $\Gamma^{LL'}_\alpha$ independent of frequency and wave vector.
	This is the generalization of the common relaxation-time approximation.
	As required, the associated transport relaxation time $\tau^L_\alpha = 1/ \Gamma^{LL}_\alpha$
differs from that found in the single-electron propagators.

The bare intraband contribution is of the Drude form
\begin{eqnarray}
\sigma^{\rm intra,0}_{\alpha \alpha} ({\bf q},\omega) 
\approx  \frac{{\rm i}  e^2 n^{\rm eff,0}_{\alpha \alpha}/m}{\omega + {\rm i} \Gamma_{1\alpha}},
\label{eq33}
\end{eqnarray}
with 
\begin{eqnarray}
n_{\alpha \alpha}^{\rm eff,0} = \frac{1}{V} \sum_{L{\bf k} \sigma}
m [v^{L,0}_\alpha ({\bf k})]^2 \bigg(-\frac{\partial f_L({\bf k})}{\partial \varepsilon_L ({\bf k})}\bigg)
\label{eq34}
\end{eqnarray}
representing the effective number of conduction electrons,
\cite{Kupcic07,Kupcic09} and $\Gamma^{LL}_\alpha = \Gamma_{1\alpha}$ by assumption.
	$n_{\alpha \alpha}^{\rm eff,0}$ includes the conduction holes from the lower band 
and the thermally activated electrons from the upper band.
	The bare interband contribution is the sum of two off-diagonal terms in (\ref{eq32}), 
with $\eta$ replaced by $\Gamma^{LL'}_\alpha$: $\sigma^{\rm inter,0}_{\alpha \alpha} ({\bf q},\omega) 
= \sum_{L \neq L'} \sigma^{LL'}_{\alpha \alpha} ({\bf q},\omega)$.

The main disadvantage of this simple procedure of calculating $\sigma_{\alpha \alpha} ({\bf q},\omega)$ is that it does not apply to low-dimensional systems with sizeable interactions.
	The reason for that is the following.
	First, comparison with experimental data shows that the relaxation rates $\Gamma^{LL'}_\alpha$
are frequency dependent.
	Second, the causality principle requires that $\Gamma^{LL'}_\alpha (\omega)$ is just the
imaginary part of a complex quantity which will be called here the electron-hole self-energy function.
	Finally, it turns out that in the strong coupling limit the electron-hole self-energy is not 
a simple function of frequency $\omega$.
	Instead, it depends quite drastically on two internal electron frequencies too.

We shall therefore present now a systematic treatment of scattering processes in interacting electron-boson systems with multiple electronic bands which treats all elements in the electron-hole self-energy consistently with the charge continuity equation (\ref{eq29}), with the causality principle and with the definition of 
${\bf E} ({\bf q}, \omega)$.
	The resulting self-consistent equations turn out to be the generalization of the equations
used by Vollhardt and W\"{o}lfle \cite{Vollhardt80} to study the scattering by static disorder in a single-band case.

\section{Bethe--Salpeter equations}

 \begin{figure}[tb]
   \centerline{\includegraphics[width=19pc]{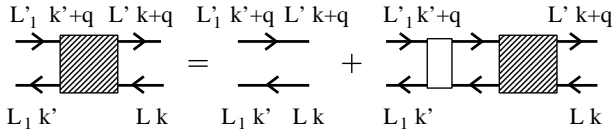}}
   \caption{The general form of the Bethe--Salpeter equations for electron-hole propagators.
   The white rectangle represents again the completely irreducible four-point interaction.
}
   \end{figure}

The term Bethe--Salpeter equations 
\cite{Negele88,Ziman88}
refers to the self-consistent equations for the exact electron-hole propagators 
$\Phi^{LL'L'_1L_1}({\bf k}, {\bf k}_+, {\bf k}'_+,{\bf k}',
{\rm i} \omega_n, {\rm i} \omega_{n+}, {\rm i} \omega_{m+}, {\rm i} \omega_m)$ shown in figure~7.
	A widely-used simplification in these equations is illustrated in figure~3.
	In this case, we have both the ladder contributions and the RPA-like contributions to 
the irreducible four-point interaction, but the three-point vertex renormalizations are neglected both in these equations and in the related Dyson equations.
	The interaction lines and the electron propagators are renormalized, so that these
equations must be treated consistently with the Dyson equations, as already pointed out in section 2.
	However, in strictly 1D metallic systems as well as in strongly-correlated low-dimensional
systems, it is necessary to use the Bethe--Salpeter equations beyond this approximation by taking 
into account the three-point vertex renormalizations as well.

To simplify the analysis, it is convenient to multiply the left-hand side of the diagrams in figure~7 by the corresponding bare vertex function and then sum over ${\bf k}'$ and ${\rm i} \omega_m$.
	For the electron-hole propagators in the dynamical conductivity (\ref{eq5}), the bare 
vertex in question is the bare dipole vertex function, and the Bethe--Salpeter equations from figure~7 turn into the Bethe--Salpeter equations for the auxiliary electron-hole propagators  
\begin{eqnarray}
&& \hspace{-10mm} 
 \Phi^{LL'}_\alpha ({\bf k}, {\bf k}_+, {\rm i} \omega_n, {\rm i} \omega_{n+})=
\frac{1}{\hbar^2} {\cal G}_{L} ({\bf k}, {\rm i} \omega_n )
\nonumber \\
&& \hspace{0mm}
\times  {\cal G}_{L'} ({\bf k}_+, {\rm i} \omega_{n+})  
(-) \Gamma^{LL'}_{\alpha}({\bf k}, {\bf k}_+, {\rm i} \omega_n, {\rm i} \omega_{n+}).
\label{eq35}
\end{eqnarray}
	The equations are illustrated in figure~8.
	
With little loss of generality, we restrict the analysis to the case where the irreducible
four-point interaction  
$\sum_\nu U_\nu^{L'L_1',L_1L} ({\bf k}_+, {\bf k}',{\bf q}', 
{\rm i} \omega_{n+}, {\rm i} \omega_{m}, {\rm i} \nu_{m})$ can be approximated by 
$U^{L'L_1',L_1L} ({\bf q}', {\rm i} \nu_{m})$.
	In this case, the explicit form of the intraband ($L=L'$) and interband ($L \neq L'$)
equations is the following 
\begin{eqnarray}
&& \hspace{-5mm}
D_{LL'}^{-1} ({\bf k},{\bf k}_+, {\rm i} \omega_n, {\rm i} \omega_{n+} ) 
\Phi^{LL'}_\alpha ({\bf k},{\bf k}_+, {\rm i} \omega_n, {\rm i} \omega_{n+} ) 
\nonumber \\
&&  \hspace{10mm}
= \frac{1}{\hbar^2} \big[ {\cal G}_{L} ({\bf k}, {\rm i} \omega_n ) - {\cal G}_{L'} ({\bf k}_+, {\rm i} \omega_{n+} ) \big] 
\nonumber \\
&&  \hspace{0mm} \times 
\bigg\{ -P_{\alpha}^{L'L}({\bf k}_+,{\bf k}) - \frac{1}{\beta} 
\sum_{L_1L_1'} \sum_{{\bf q}' {\rm i} \nu_m} U^{L'L_1',L_1L} ({\bf q}', {\rm i} \nu_{m})
\nonumber \\
&&  \hspace{10mm}
\times \Phi^{L_1L_1'}_\alpha ({\bf k}+{\bf q}',{\bf k}_++{\bf q}', {\rm i} \omega_m, {\rm i} \omega_{m+} ) \bigg\},
\label{eq36}
\end{eqnarray}
with
\begin{eqnarray}
&&  \hspace{-10mm}
D_{LL'}^{-1}({\bf k},{\bf k}_+, {\rm i} \omega_n, {\rm i} \omega_{n+}) 
=  {\rm i} \nu_{n} + \varepsilon_L({\bf k})/\hbar -\varepsilon_{L'}({\bf k}_+)/\hbar
\nonumber \\
&& \hspace{20mm}  
+ \Sigma_L ({\bf k}, {\rm i} \omega_n ) - \Sigma_{L'} ({\bf k}_+, {\rm i} \omega_{n+})
\label{eq37}
\end{eqnarray}
and ${\rm i} \omega_m = {\rm i} \omega_n + {\rm i} \nu_m$.
	The ${\cal G}_{L} ({\bf k}, {\rm i} \omega_n)$ satisfy the Dyson equations for electrons
(\ref{eq1}) and $U^{L'L_1',L_1L} ({\bf q}', {\rm i} \nu_{m})$ is a simple generalization of the completely irreducible four-point interaction $U^{L'L,LL'} ({\bf q}', {\rm i} \nu_{m})$ from (\ref{eq21}).

\begin{figure}
   \centerline{\includegraphics[width=19pc]{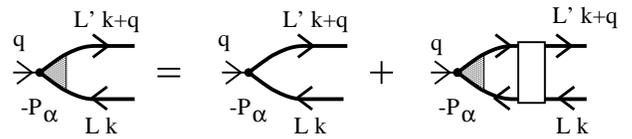}}
   \caption{The Bethe-Salpeter equations (\ref{eq36}) for auxiliary electron-hole propagators.
   }
  \end{figure}

In the electronic multiband model we are interested in here, there are $n$ bands in the vicinity of the Fermi level.
	The bands are built of $n$ orbitals per unit cell.
	Associated with each of these orbitals are the intracell charge and spin fluctuations, 
and these intracell fluctuations are possible sources of scattering.
	The monopole charge fluctuations underlying the present analysis of the dynamical conductivity
tensor might be also strongly affected by the presence of the intracell fluctuations.
	As a result, $\sigma_{\alpha \beta} ({\bf q}, \omega)$ will have considerably more structure than
that characterizing the ordinary single-band model, even in the case where only one band out of $n$ valence bands intersects the Fermi level.

\subsection{Quantum transport equation}
In the remainder of this section we consider the intraband term in (\ref{eq36}) from this point of view.
	We assume that there is only one conduction band and again drop reference to the band index.
	We use the intraband scattering approximation where the scattering processes in which 
the electron changes the band are neglected.
	As discussed in Article II,
this approximation is easily seen not to be adequate, for example, to study fluctuations
in the order parameter in the pseudogap state of the ordinary CDW systems.
	However, it is convenient to illustrate the main ideas regarding the quantum transport equations.
	Finally, the three-point vertex renormalizations in the renormalized interactions are 
also neglected, as usual in the common non-crossing approximation for the electron-boson systems.
\cite{Millis87,Niksic95,Rozenberg96}
	The resulting intraband Bethe--Salpeter equation is of the form
\begin{eqnarray}
&& \hspace{-5mm}
D^{-1} ({\bf k},{\bf k}_+, {\rm i} \omega_n, {\rm i} \omega_{n+} ) 
\Phi_\alpha ({\bf k},{\bf k}_+, {\rm i} \omega_n, {\rm i} \omega_{n+} ) 
\nonumber \\
&&  \hspace{5mm}
= \frac{1}{\hbar^2} \big[ {\cal G} ({\bf k}, {\rm i} \omega_n ) - {\cal G} ({\bf k}_+, {\rm i} \omega_{n+} ) \big] 
\bigg\{ p_{\alpha}(-{\bf q}) 
\nonumber \\
&&  \hspace{-5mm}
- \frac{1}{\beta} \sum_{{\bf q}' {\rm i} \nu_m} {\cal F} ({\bf q}', {\rm i} \nu_{m})
\Phi_\alpha ({\bf k}+{\bf q}',{\bf k}_++{\bf q}', {\rm i} \omega_m, {\rm i} \omega_{m+} ) \bigg\}.
\nonumber \\
\label{eq38}
\end{eqnarray}

The structure of (\ref{eq38}) is simplified by defining the auxiliary single-electron self-energy function
\begin{eqnarray}
&& \hspace{-8mm} 
\hbar \widetilde \Sigma ({\bf k}, {\rm i} \omega_n ) = - \frac{1}{\beta \hbar} \sum_{{\bf q}' {\rm i} \nu_m}
\bigg\{1 - \frac{v^0_{\alpha}({\bf k}+{\bf q}')}{v^0_{\alpha}({\bf k})} \bigg\} 
\nonumber \\
&& \hspace{10mm}  \times {\cal F} ({\bf q}', {\rm i} \nu_{m}) 
{\cal G} ({\bf k}+{\bf q}', {\rm i} \omega_{n}+ {\rm i} \nu_m)
\label{eq39}
\end{eqnarray}
and the electron-hole self-energy
\begin{eqnarray}
\Pi ({\bf k},{\bf k}_+, {\rm i} \omega_n,{\rm i} \omega_{n+})= 
\widetilde \Sigma ({\bf k}, {\rm i} \omega_n )-
\widetilde \Sigma ({\bf k}_+, {\rm i} \omega_{n+}).
\label{eq40}
\end{eqnarray}
	Equation (\ref{eq38}) now reduces to the quantum transport equation
\begin{eqnarray}
&& \hspace{-10mm}
\big[ {\rm i} \nu_{n} + \varepsilon({\bf k})/\hbar -\varepsilon({\bf k}_+)/\hbar
\nonumber \\
&&  \hspace{-2mm}
+  \Pi^r ({\bf k},{\bf k}_+, {\rm i} \omega_n, {\rm i} \omega_{n+})  \big]
\Phi_\alpha ({\bf k},{\bf k}_+, {\rm i} \omega_n, {\rm i} \omega_{n+} ) 
\nonumber \\
&&  \hspace{5mm}
- \frac{1}{\hbar^2} \big[ {\cal G} ({\bf k}, {\rm i} \omega_n ) 
- {\cal G} ({\bf k}_+, {\rm i} \omega_{n+} ) \big] p_\alpha (-{\bf q}) 
\nonumber \\
&&  \hspace{-5mm}
=  - {\rm i} \Pi^i ({\bf k},{\bf k}_+, {\rm i} \omega_n, {\rm i} \omega_{n+})
\Phi_\alpha ({\bf k},{\bf k}_+, {\rm i} \omega_n, {\rm i} \omega_{n+} ).
\label{eq41}
\end{eqnarray}
	Equation~(\ref{eq41}) is the self-consistent equation for the auxiliary electron-hole propagator, 
in contrast to the ordinary transport equation, equation~(\ref{eq47}), which is formulated in terms of the induced density $\delta n ({\bf k}, {\bf q}, {\rm i} \nu_n)$ defined by
\begin{equation}
\delta n ({\bf k}, {\bf q}, {\rm i} \nu_n) = \frac{1}{\beta}\sum_{{\rm i} \omega_n } 
\Phi_{\alpha} ({\bf k},{\bf k}_+,  {\rm i} \omega_n, {\rm i} \omega_{n+}) E_\alpha ({\bf q},  {\rm i} \nu_n).
\label{eq42}
\end{equation}
	The conditions under which (\ref{eq41}) reduces to the ordinary transport equation are 
discussed below.

To illustrate the role  of the electron-EM field vertex renomalizations in (\ref{eq41}), it is useful to define the auxiliary single-electron propagator $\widetilde {\cal G} ({\bf k}, {\rm i} \omega_n )$ as the single-particle propagator which satisfies the following Dyson equation
\begin{eqnarray}
\big[{\rm i} \hbar \omega_n - \varepsilon ({\bf k}) + \mu 
-\hbar \widetilde \Sigma ({\bf k}, {\rm i} \omega_n) \big]
\widetilde {\cal G} ({\bf k}, {\rm i} \omega_n) = \hbar.
\label{eq43}
\end{eqnarray}
	The forward scattering processes cancel identically out in 
$\widetilde \Sigma ({\bf k}, {\rm i} \omega_n)$  as a result of the charge continuity equation.
	The fact that the auxiliary propagator 
$\widetilde {\cal G} ({\bf k}, {\rm i} \omega_n)$ has the quasi-particle pole even for relatively strong scattering processes is consistent with the fact that the residue associated with 
the quasi-particle pole in the exact propagator ${\cal G} ({\bf k}, {\rm i} \omega_n)$ is proportional to $t_b/t_a$ and vanishes in the strictly 1D limit $t_b \rightarrow 0$. \cite{Dzyaloshinskii73}
	This interesting observation is in the background of a variety of approximate
treatments of various pseudogapped states.
	In these approaches, the forward scattering contributions to the FFCF 
${\cal F} ({\bf q}', {\rm i} \nu_{m})$ are neglected from the outset.

The expression for the intraband conductivity which correctly describes the $\omega \approx 0$ 
Drude limit, but which is not necessarily correct in the $\omega=0$ Thomas--Fermi limit, is the following
\begin{eqnarray}
&& \hspace{-12mm}  
\sigma_{\alpha \alpha}^{\rm intra} ({\bf q}, {\rm i} \nu_n) =  \frac{1}{V}  \sum_{{\bf k} \sigma}  
J_{\alpha}({\bf k}) p_{\alpha}(-{\bf q}) \frac{1}{\beta \hbar} \sum_{{\rm i} \omega_n} 
 \nonumber \\
 && \hspace{-7mm}  
\times \frac{ {\cal G} ({\bf k}, {\rm i} \omega_n ) - {\cal G} ({\bf k}_+, {\rm i} \omega_{n+} )  }{
{\rm i}  \hbar \nu_n + \hbar \Pi ({\bf k},{\bf k}_+, {\rm i} \omega_n, {\rm i} \omega_{n+})  + \varepsilon({\bf k}) - \varepsilon({\bf k}_+)}.
\label{eq44}
\end{eqnarray}
	This is nothing but (\ref{eq32}) for the ideal intraband conductivity in which ${\rm i} \eta$ 
is replaced by $\Pi ({\bf k},{\bf k}_+, {\rm i} \omega_n,{\rm i} \omega_{n+})$.

\subsection{Memory-function approximation}
In the weak coupling limit, the self-energy $\Pi ({\bf k},{\bf k}_+, {\rm i} \omega_n,{\rm i} \omega_{n+})$ is 
a non-singular function of two fermion frequencies and in the simplest approximation
can be written in the form 
$\Pi ({\bf k},{\bf k}_+, {\rm i} \omega_n,{\rm i} \omega_{n+}) \approx M_\alpha({\bf k}, {\rm i} \nu_n) 
= M^r_\alpha({\bf k}, {\rm i} \nu_n) + {\rm i} M^i_\alpha({\bf k}, {\rm i} \nu_n)$.
	$M_\alpha({\bf k}, {\rm i} \nu_n)$ is called the memory function.
\cite{Gotze72,Giamarchi91,Kupcic04}
	The momentum distribution function
\begin{equation}
n({\bf k}) = \frac{1}{\beta \hbar} \sum_{{\rm i} \omega_n} {\cal G} ({\bf k}, {\rm i} \omega_n ) 
{\rm e} ^{{\rm i} \omega_n \eta}
\label{eq45}
\end{equation}
appearing in (\ref{eq44}) in the memory-function approximation
reduces in the weak coupling limit to the Fermi--Dirac distribution function $f({\bf k})$.

We now define the electron-boson coupling function $\lambda_\alpha({\bf k},\omega)$, the mass enhancement factor $m_{\alpha \alpha}^*({\bf k}, \omega)/m$, the intraband relaxation rate 
$\Gamma_{1\alpha} ({\bf k}, \omega)$ and the collision integral $I({\bf k}, \omega)$ by
\begin{eqnarray}
&& \hspace{-10.5mm}
\lambda_\alpha({\bf k},\omega) = M^r_\alpha ({\bf k}, \omega)/\omega,
\nonumber \\
&& \hspace{-13.5mm} 
m_{\alpha \alpha}^*({\bf k}, \omega) = (1+ \lambda_\alpha({\bf k},\omega))m,
\nonumber \\
&& \hspace{-12mm}
\Gamma_{1 \alpha} ({\bf k}, \omega) = (m/m_{\alpha \alpha}^*({\bf k}, \omega)) 
M^i_\alpha ({\bf k}, \omega),
\nonumber \\
&& \hspace{-8mm} 
I({\bf k}, \omega) = - \Gamma_{1\alpha} ({\bf k}, \omega) \delta n ({\bf k}, {\bf q}, \omega).
\label{eq46}
\end{eqnarray}
	Summation of (\ref{eq41}) over ${\rm i} \omega_n$, together with  
analytical continuation ${\rm i} \nu_n \rightarrow \omega + {\rm i} \eta$ and multiplication by $E_\alpha ({\bf q}, \omega)$, gives the ordinary transport equation
\cite{Pines89,Ziman72,Abrikosov88}
\begin{eqnarray}
&& \hspace{-5mm}
-{\rm i} \big[\hbar \omega + E({\bf k}) - E({\bf k}_+) \big] \delta n ({\bf k}, {\bf q}, \omega)
+ \frac{m}{m_{\alpha \alpha}^*({\bf k}, \omega)} e \hbar v_\alpha^0 ({\bf k}) 
\nonumber \\
&&  \hspace{20mm}
\times \frac{\partial f({\bf k})}{\partial \varepsilon ({\bf k})} E_\alpha ({\bf q}, \omega)
=  \hbar I({\bf k}, \omega),
\label{eq47}
\end{eqnarray}
where $E({\bf k}) = \varepsilon({\bf k})/(1 + \lambda_\alpha({\bf k},\omega=0))$.

In solving (\ref{eq47}) it is convenient to follow the usual textbook procedure. \cite{Pines89}
	This equation can be solved either for $J_\alpha^{\rm c} ({\bf q},\omega)$ or for 
$\rho^{\rm c} ({\bf q},\omega)$, where
\begin{eqnarray}
&& \hspace{-10mm}
\rho^{\rm c} ({\bf q},\omega) = \frac{1}{V} \sum_{{\bf k}\sigma}  e \delta n_0({\bf k}, {\bf q},\omega),
\nonumber \\
&& \hspace{-11mm}
J_\alpha^{\rm c} ({\bf q},\omega) = 
\frac{1}{V} \sum_{{\bf k}\sigma} e v_\alpha^0 ({\bf k}) \delta n_1({\bf k}, {\bf q},\omega)
\label{eq48}
\end{eqnarray}
are, respectively, the induced charge and current densities of conduction electrons,
and the $\delta n_i({\bf k}, {\bf q},\omega)$, $i = 0,1$, are the contributions to 
$\delta n({\bf k}, {\bf q},\omega)$ which are proportional to $[v^0_\alpha ({\bf k})]^i$.
	For further considerations of the random-phase approximation in section 8.1, 
the second route is more appropriate choice.
	In this case, we consider the thin slab with the longitudinal polarization of the electric
fields, introduce the scalar fields $V^{\rm tot} ({\bf q},\omega)$ and 
$V^{\rm ext} ({\bf q},\omega)$,
by the relations $V^{\rm tot} ({\bf q},\omega) = ({\rm i}/q_\alpha) E_\alpha ({\bf q}, \omega)$ 
and $V^{\rm ext} ({\bf q},\omega) = ({\rm i}/q_\alpha) E_{0\alpha} ({\bf q}, \omega)$,
and use the definition of the dielectric susceptibility 
$\chi^{\rm intra} ({\bf q},\omega)$: $\rho^{\rm c} ({\bf q},\omega) 
= \chi^{\rm intra} ({\bf q},\omega) V^{\rm tot} ({\bf q},\omega)$, or 
\begin{equation}
[1/ \chi^{\rm intra} ({\bf q},\omega) - 4\pi/q^2] \rho^{\rm c} ({\bf q},\omega) 
= V^{\rm ext} ({\bf q},\omega).
\label{eq49}
\end{equation}
	It is readily seen that the intraband conductivity is of the form
\begin{eqnarray}
&& \hspace{-10mm}
\sigma^{\rm intra}_{\alpha \alpha} ({\bf q},\omega) \equiv \frac{{\rm i} \omega}{q_\alpha^2} 
\chi^{\rm intra} ({\bf q},\omega)
= {\rm i} e^2 \frac{1}{V} \sum_{{\bf k} \sigma} \frac{m}{m_{\alpha \alpha}^*({\bf k}, \omega)}
\nonumber \\
&& \hspace{10mm}
\times \frac{[v^0_\alpha ({\bf k})]^2 \big( -\partial f({\bf k})/\partial \varepsilon ({\bf k})\big)}{
\omega + {\rm i} \Gamma_{1\alpha}({\bf k},\omega) -q_\alpha^2 v^2_\alpha ({\bf k})/\omega}.
\label{eq50}
\end{eqnarray}
	In the Drude $\omega \approx 0$ limit, we set $q_\alpha$ in the denominator equal to zero, with
$\Gamma_{1\alpha}({\bf k},\omega) \approx \langle \Gamma_{1 \alpha} ({\bf k}, \omega) \rangle_{\rm FS}=
\Gamma_{1 \alpha} (\omega)$ and $m_{\alpha \alpha}^* ({\bf k},\omega) \approx  
\langle m_{\alpha \alpha}^*({\bf k}, \omega) \rangle_{\rm FS} = m_{\alpha \alpha}^*(\omega)$, to obtain 
\begin{eqnarray}
\sigma^{\rm intra}_{\alpha \alpha} ({\bf q}, \omega) = 
\frac{{\rm i} e^2n_{\alpha \alpha}^{\rm eff,0}/m_{\alpha \alpha}^*(\omega) }{\omega + {\rm i} \Gamma_{1\alpha}(\omega)},
\label{eq51}
\end{eqnarray}
an expression known as the generalized Drude formula.
	Here, $\langle \ldots \rangle_{\rm FS}$ means the  average over the Fermi surface.
	The expression (\ref{eq51}) remains valid when more than one band intersects the Fermi 
level, provided that $\Gamma^{LL}_{\alpha}(\omega) \approx \Gamma_{1\alpha}(\omega)$.
	The generalization to the case with the relaxation rates $\Gamma^{LL}_{\alpha}(\omega)$ dependent 
on the band index $L$ is straightforward.

Although we have arrived at (\ref{eq41}) and (\ref{eq47}) by considering in
the auxiliary self-energy $\widetilde \Sigma ({\bf k}, {\rm i} \omega_n )$ only the scattering by boson modes, these expressions are of a more general validity, namely all relevant scattering mechanisms 
have to be included in $\Pi ({\bf k},{\bf k}_+, {\rm i} \omega_n, {\rm i} \omega_{n+} )$.
	Therefore, we can use (\ref{eq41}) to study the effects of local 
and short-range electron-electron interactions on electrodynamic properties of strongly-correlated low-dimensional systems or to study similar effects associated with the scattering by different periodic CDW/SDW potentials.

\section{Interband conductivity}
In analogy to (\ref{eq44}), we can write the bare interband conductivity in the form
\begin{eqnarray}
&& \hspace{-5mm}  
\sigma_{\alpha \alpha}^{\rm inter,0} ({\bf q}, {\rm i} \nu_n) 
=  -\frac{1}{V} \sum_{L \neq L'} \sum_{{\bf k} \sigma} \frac{1}{\beta \hbar} \sum_{{\rm i}\omega_n}
J_{\alpha}^{LL'}({\bf k}) P_{\alpha}^{L'L}({\bf k})
\nonumber \\
 && \hspace{0mm}  
\times
\frac{ {\cal G}_{L} ({\bf k}, {\rm i} \omega_n ) - {\cal G}_{L'} ({\bf k}_+, {\rm i} \omega_{n+})}{
{\rm i} \hbar \nu_n + \hbar \Pi_{LL'} ({\bf k},{\bf k}_+, {\rm i} \omega_n, {\rm i} \omega_{n+})  
 - \varepsilon_{L'}({\bf k}_+) + \varepsilon_{L}({\bf k})},
\nonumber \\
\label{eq52}
\end{eqnarray}
with the electron-EM field vertex corrections neglected, resulting in 
$\Pi ({\bf k},{\bf k}_+, {\rm i} \omega_n,{\rm i} \omega_{n+}) 
\approx \Sigma ({\bf k}, {\rm i} \omega_n )- \Sigma ({\bf k}_+, {\rm i} \omega_{n+})$.
	The electron-hole self-energy $\Pi_{LL'} ({\bf k},{\bf k}_+, {\rm i} \omega_n,{\rm i} \omega_{n+})$ 
is replaced by $M_\alpha^{LL'} ({\bf k}, {\rm i} \nu_n)$, in the memory-function approximation, and by 
${\rm i} \Gamma_\alpha^{LL'} ({\bf k}) \approx {\rm i} \Gamma_\alpha^{LL'}$, in the relaxation-time approximation.

In those multiband systems in which the interband dipole vertex $P_{\alpha}^{L'L}({\bf k})$ is purely imaginary, the local dipole moment vanishes and there are no local field effects.
\cite{Kupcic02}
	The interband conductivity is given by $\sigma_{\alpha \alpha}^{\rm inter} ({\bf q}, \omega)
=\sigma_{\alpha \alpha}^{\rm inter,0} ({\bf q}, \omega)$.
	This result can be equally well understood to represent the solution to the RPA equation
\begin{equation}
[1/ \chi ({\bf q},\omega) - 4\pi/q^2] \rho ({\bf q},\omega) = V^{\rm ext} ({\bf q},\omega)
\label{eq53}
\end{equation}
for the total dielectric susceptibility
$ \chi ({\bf q},\omega) =  \sum_{i={\rm intra},{\rm inter}}\chi^{i} ({\bf q},\omega)$, 
where $\chi^{i} ({\bf q},\omega) = (q_\alpha^2/{\rm i} \omega) \sigma^{i}_{\alpha \alpha} ({\bf q},\omega)$.

In the opposite case, the interband dipole vertex is real and its dispersionless part is 
directly related to the intracell dipole transitions.
	Particularly interesting are the examples in which
$P_{\alpha}^{L'L}({\bf k}_+,{\bf k}) \approx P_{\alpha}^{L'L}$ is 
completely local. 
	In this case we can use the procedure from the following section to determine the effects of 
the Lorentz local fields.
	
At this point caution is in order regarding the relaxation-time approximation.
	It is well known that in those systems in which the damping energy $\Gamma_\alpha^{LL'} ({\bf k})$
is comparable to the interband threshold energy (the pseudogapped state of the ordinary CDW systems is an example) the relaxation-time approximation of (\ref{eq52}) overestimates the in-gap part of the interband conductivity.
	In this regime, $\sigma_{\alpha \alpha}^{\rm inter,0} ({\bf q}, \omega)$ is found to be 
better described by the approximate textbook expression
\cite{Wooten72,Ziman79}
\begin{eqnarray}  
&& \hspace{-12mm}
\sigma^{\rm inter,0}_{\alpha \alpha} ( \omega) 
\nonumber \\
&& \hspace{-8mm}
= \frac{1}{V} \sum_{L \neq L'} \sum_{{\bf k} \sigma} 
\frac{{\rm i} \omega | P_{\alpha}^{LL'}({\bf k})|^2 [f_{L}({\bf k})- f_{L'}({\bf k}_+)]}{
\hbar \omega + {\rm i} \hbar \Gamma_{2\alpha}({\bf k}) - \varepsilon_{L'}({\bf k}_+) 
+ \varepsilon_{L}({\bf k})}.
\label{eq54}
\end{eqnarray}

In order to briefly illustrate our results, the typical mean-field single-particle conductivity of the Peierls CDW model at temperatures well below the mean-field transition temperature $T_{\rm P}^0$ is shown in figure 9.
	The dashed and the dot-dashed lines represent, respectively, the interband contributions calculated
by using (\ref{eq54}) and the relaxation-time  version of (\ref{eq52}).
	Notice the difference between the two curves at $\omega \approx 0$.
	The solid line is the total single-particle conductivity which is the sum of the contribution 
of thermally activated charge carriers in (\ref{eq51}) and the interband contribution (\ref{eq54}).
	This result agrees reasonably well with the experimental spectra \cite{Degiorgi91}
measured at temperatures well below the CDW transition temperature.
	The detailed discussion of the mean-field approximation as well as of the effects of fluctuations in 
the order parameter is given in Article II.

\begin{figure}[tb]
  \centerline{\includegraphics[width=15pc]{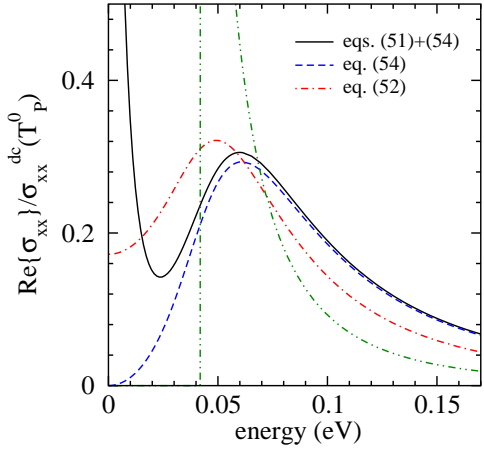}}
  \caption{
  The mean-field single-particle conductivity of the Peierls CDW model 
  along the highly conducting direction $x$, at temperature $T = 160$ K.
  The bare electron dispersion is 
  $\varepsilon ({\bf k}) = - \sum_{\alpha=x,y} 2t_\alpha \cos k_\alpha a_\alpha$, 
  with $2t_a = 0.5$ eV and $2t_b = 50$ meV.
  The mean-field transition temperature is assumed to be $T_{\rm P}^0 = 200$ K and 
  $\Delta (160\, {\rm K}) =21$ meV is the CDW order parameter at temperature $T= 160$ K.
  The damping energies are $5 \hbar \Gamma_1 = \hbar \Gamma_2 = 30$ meV 
  at $T < T_{\rm P}^0$, and $\hbar \Gamma^0_1 = 30$ meV at $T > T_{\rm P}^0$.
  The dot-dot-dashed line shows the inverse square-root singularity in
  $\sigma^{\rm inter, sp}_{xx} (\omega)$ associated with the limit $\hbar \Gamma_2 \rightarrow 0$.
  }
  \end{figure}

In the next section, we shall examine the Lorentz local field corrections in the relaxation-time approximation, starting with $\sigma^{\rm inter, 0}_{\alpha \alpha} ({\bf q},\omega)$  given by (\ref{eq54}). 
	The damping energy $\hbar \Gamma_{2 \alpha}({\bf k}) \approx \hbar \Gamma_{2 \alpha}$ is assumed 
to be small in comparison with the interband threshold energy.

\section{Local field effects}
Most Q1D systems are characterized by a large number of atoms in the unit cell (for example, in the blue bronze K$_{0.3}$MoO$_3$ we have 86 atoms per unit cell).
	The bands in the vicinity of the Fermi level are thus built of different molecular orbitals 
of the well-defined symmetry with the intermolecular distance large when compared to the first-neighbor atom-atom distance.
	Therefore, the bands are narrow and the typical interband threshold energy is 
in the infrared part of the spectrum.
	In most cases of interest, the low-energy interband electron-hole excitations are local 
and correspond to dipolar transitions between one molecular orbital of the $s$ symmetry and another of the $p_\alpha$ symmetry.
	Consequently, to determine the proper structure of the total conductivity tensor 
$\sigma_{\alpha \alpha} ({\bf q}, \omega)$ in this case, it is important to take into account the local field effects too.

This is not the only Q1D case in which the local field corrections are important.
	In the ordered state of the strong bond-CDW systems the interband 
dipole excitations are also local.
	Finally, the ordinary $sp_\alpha$ models with the interband threshold energy in the visible part 
of the spectrum also have $\sigma^{\rm inter,0}_{\alpha \alpha} ({\bf q}, \omega)$ of the form (\ref{eq54}), with the dipole vertex $P_{\alpha}^{L'L}({\bf k}) \approx P_{\alpha}^{L'L}$ real. \cite{Wooten72,Adler62,Wieser62}

\subsection{Multicomponent random-phase approximation}
To demonstrate the usual procedure for dealing with the local field corrections, 
\cite{Adler62,Wieser62,Zupanovic96} 
we consider the model with the intraband currents and with one type of bound charges.
	We assume that the contribution of the bound charges is related to the intra-atomic 
or intramolecular dipolar transitions in $\sigma^{\rm inter,0}_{\alpha \alpha} ({\bf q}, \omega)$.
	In section 8.2, we shall briefly discuss the case of infrared-active optical phonons, as well.
	In this section, we make a simple modification of the notation.
	For example, the label intra is replaced by c and the label inter by b
(for conduction and bound charges).

The bare intraband dielectric susceptibility is assumed to be of the form
\begin{eqnarray}
\chi^{\rm c,0} ({\bf q},\omega) \equiv \chi^{0}_{11} ({\bf q},\omega)
= (q_\alpha^2/{\rm i}\omega) \sigma^{\rm c,0}_{\alpha \alpha} ({\bf q},\omega),
\label{eq55}
\end{eqnarray}
with $\sigma^{\rm c,0}_{\alpha \alpha} ({\bf q},\omega)$ given by (\ref{eq50}), and 
the bare interband dielectric susceptibility is
\begin{eqnarray}
\chi^{\rm b,0} ({\bf q},\omega) \equiv \chi^{0}_{22} ({\bf q},\omega)
= (q_\alpha^2/{\rm i}\omega) \sigma^{\rm b,0}_{\alpha \alpha} ({\bf q},\omega),
\label{eq56}
\end{eqnarray}
with $\sigma^{\rm b,0}_{\alpha \alpha} ({\bf q},\omega)$ given by (\ref{eq54}).

   \begin{figure}[tb]
   \centerline{\includegraphics[width=20pc]{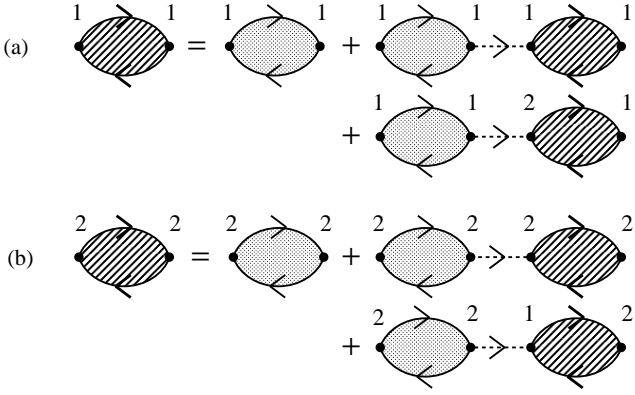}}
   \caption{The RPA equations for the screened susceptibilities $\chi_{11} ({\bf q}, \omega)$ (a) and 
    $\chi_{22} ({\bf q}, \omega)$ (b).
   }
   \end{figure}

The electrons are assumed to be subjected to an external longitudinal electric field ${\bf E}_0 ({\bf r},t)$ represented by the scalar potential $V^{\rm ext} ({\bf r},t)$.
	$\rho_1 ({\bf q}, \omega) = \rho^{\rm c} ({\bf q}, \omega)$ and 
$\rho_2 ({\bf q}, \omega) = -{\rm i} q_\alpha P_\alpha^{\rm b} ({\bf q}, \omega)$ are the Fourier transforms of two types of macroscopic charge densities.
	$\rho^{\rm c} ({\bf q}, \omega)$ describes the monopole-charge density of conduction
electrons and $P_\alpha^{\rm b} ({\bf q}, \omega)$ the dipole-charge density of bound charges.
	The two-component random-phase approximation describes the long-range screening in such 
a case.
	According to figure~10, the RPA equations can be shown in the form
\begin{eqnarray}
&& \hspace{-5mm}
\big[1/\chi_{11}^0 ({\bf q}, \omega) - V_{11}\big] \rho_1 ({\bf q}, \omega) 
- V_{12} \, \rho_2 ({\bf q}, \omega) = V^{\rm ext} ({\bf q},\omega) ,
\nonumber \\
&& \hspace{-7mm}
- V_{21} \, \rho_1 ({\bf q}, \omega) + \big[1/\chi_{22}^0 ({\bf q}, \omega) 
- V_{22}\big] \rho_2 ({\bf q}, \omega) = V^{\rm ext} ({\bf q},\omega).
\nonumber \\
\label{eq57}
\end{eqnarray}
	Here, the $\chi_{ii}^0 ({\bf q}, \omega)$ stand for two bare dielectric susceptibilities 
in (\ref{eq55}) and (\ref{eq56}), and $V_{ij} \equiv V_{ij}({\bf q}) = - N_{ij}/q^2$ are the Fourier transforms of different long-range interactions.
	The numbers $N_{11} = -4 \pi$, $N_{12} = N_{21} =-4 \pi$ and $N_{22} = -4 \pi + 4 \pi /3$ 
correspond to the monopole-monopole, monopole-dipole and dipole-dipole interactions, respectively. \cite{Zupanovic96}
	$N_{11} = N_{12} = N_{21} =-4 \pi$ are easily shown to represent the common depolarization
factors for the longitudinal ${\bf q} \approx 0$ fields. \cite{Landau95}
	It is not hard to verify that,
after neglecting the Lorentz term $4 \pi/3$ in $N_{22}$ 
(i.e., when $N_{22} = N_{11}$), equations~(\ref{eq57}) reduce to (\ref{eq53}).

Equations (\ref{eq57}) may be rewritten now in the alternative form, in terms of two dipole densities 
$P_{i \alpha} ({\bf q}, \omega) = ({\rm i}/q_\alpha) \rho_i ({\bf q}, \omega)$, to give
\begin{eqnarray}
&& \hspace{15mm}
\big( 1/\alpha_{11}^0 \big) P_{1 \alpha}   = E_{0 \alpha} + N_{11} P_{1 \alpha}  + N_{12} P_{2 \alpha},
\nonumber \\
&& \hspace{-5mm}
\big( 1/\alpha_{22}^0   + N_{11} - N_{22} \big) P_{2 \alpha} 
= E_{0 \alpha}  + N_{21} P_{1 \alpha} + N_{11} P_{2 \alpha},
\nonumber \\
\label{eq58}
\end{eqnarray}
with the implicit dependence on ${\bf q}$ and  $\omega$.
	Here
\begin{eqnarray}
\alpha^0_{ii}({\bf q}, \omega) = (-1/q_\alpha^2)\chi^0_{ii}({\bf q}, \omega)
\label{eq59}
\end{eqnarray}
is the common notation for two bare polarizabilities.
	It must be emphasized that the 
expressions on the right-hand side of (\ref{eq58}) represent the macroscopic 
electric field ${\bf E} ({\bf q}, \omega)$, equation~(\ref{eq30}), 
which is the sum of the external electric field and two depolarization fields originating from the conduction electrons and from the induced dipole moments of bound charges. \cite{Landau95}
	This is a useful observation which helps us to rewrite (\ref{eq58}) in the final form
\begin{eqnarray}
\big[ 1/\sigma^{\rm c}_{\alpha \alpha} ({\bf q}, \omega) \big] J^{\rm c}_{\alpha} ({\bf q}, \omega) =
E_{\alpha}({\bf q}, \omega),
\label{eq60} \\
\big[ 1/\alpha^{\rm b}_{\alpha \alpha} ({\bf q}, \omega) \big] P^{\rm b}_{\alpha} ({\bf q}, \omega) =
E_{\alpha}({\bf q}, \omega),
\label{eq61} 
\end{eqnarray}
with
\begin{equation}
E_{\alpha}({\bf q}, \omega) = E_{0 \alpha}({\bf q}, \omega) 
- 4 \pi \big[ P_{1 \alpha}({\bf q}, \omega)  +  P_{2 \alpha}({\bf q}, \omega) \big].
\label{eq62} 
\end{equation}
	For the thin slab with the longitudinal polarization of the fields, 
the field ${\bf E}_0 ({\bf r},t)$ is identical
to the dielectric displacement field ${\bf D} ({\bf r},t)$ and the relation (\ref{eq62}) leads to the usual form of the longitudinal macroscopic dielectric function.
	If other high-energy on-site dipolar excitations are also included, 
the total dielectric function becomes
\begin{equation}
\varepsilon_{\alpha \alpha} ({\bf q}, \omega) = \varepsilon^{\rm HE} _{\alpha \alpha} ({\bf q}, \omega)
+ \frac{4 \pi {\rm i} \sigma^{\rm c}_{\alpha \alpha} ({\bf q}, \omega)}{\omega} 
+ 4 \pi \alpha^{\rm b}_{\alpha \alpha} ({\bf q}, \omega).
\label{eq63} 
\end{equation}
	Here,
$\alpha_{11}({\bf q}, \omega) = \alpha^0_{11}({\bf q}, \omega) \equiv ({\rm i}/\omega)
\sigma^{\rm c}_{\alpha \alpha} ({\bf q}, \omega)$ and
\begin{equation}
\alpha_{22}({\bf q}, \omega) 
= \frac{\alpha^0_{22}({\bf q}, \omega)}{1 - (4\pi/3) \alpha^0_{22}({\bf q}, \omega)}
\equiv \alpha^{\rm b}_{\alpha \alpha} ({\bf q}, \omega)
\label{eq64} 
\end{equation}
are two renormalized polarizabilities.

In the notation of the previous sections, this result can be expressed by (\ref{eq10}) and (\ref{eq11}).
	We can therefore conclude that the contribution of the monopole charges to the conductivity tensor 
is decoupled  from the contribution of the dipole charges.
	The latter contribution is affected by the Lorentz field in the usual way, 
giving also the possibility for the collective contribution in 
$\sigma^{\rm inter}_{\alpha \alpha} ({\bf q},\omega)$ when
$1 + (4\pi /3 \omega) {\rm Im} \{\sigma^{\rm inter,0}_{\alpha \alpha} ({\bf q},\omega)\} = 0$.

\subsection{Infrared-active optical phonons}

It was implied by our discussion of the quantum transport equation that (\ref{eq5}),
together with the density operators (\ref{eq3}) and (\ref{eq23}), is able to capture all relevant aspects of the electron-phonon coupling in the dynamical conductivity tensor of multiband electronic systems, at least for the scattering by acoustic and Raman-active optical phonons.
	In order to treat carefully infrared-active optical phonons, we must also include
the direct coupling between phonons and EM fields
in $H^{\rm ext}$ and take into account the corresponding local field corrections.

The generalization of (\ref{eq63}) to such a case with multiple dipolar bound-charge contributions is very simple. 
	For our present purposes it is simplest to replace $\alpha^0_{22}({\bf q}, \omega)$ by 
the sum of all bare dipolar bound-charge polarizabilities.
	In the two-band case with one infrared-active phonon branch, we obtain 
\begin{eqnarray}
\alpha^0_{22}({\bf q}, \omega) = (-1/q_\alpha^2)\chi^{\rm inter, 0}({\bf q}, \omega)
+ \alpha^0_{\rm ph}({\bf q}, \omega),
\label{eq65}
\end{eqnarray}
with $\alpha^0_{\rm ph}({\bf q}, \omega) = 
(e^2/M_{\nu} V_0) [\omega_{\nu {\bf q}}^2 - \omega(\omega + {\rm i} \gamma)]^{-1}$
being the bare phonon polarizability.

\section{Conclusions}
In this article, we have derived the dynamical conductivity tensor in a general multiband electronic model with sizeable boson-mediated electron-electron interactions.
	Starting with the gauge-invariant form of the coupling between valence electrons 
and external EM fields, we have derived the general form of
the self-consistent Bethe--Salpeter equations for the auxiliary intraband and interband electron-hole propagators.
	It is demonstrated that the intraband Bethe--Salpeter equation is identical to the 
usual quantum transport equation and that it reduces to the ordinary transport equation in the weak coupling limit.
	In this way, the present response theory of multiband electronic systems is connected 
to the generalized Drude formula for the intraband conductivity widely used in experimental analyses as well as to different Fermi liquid theories based on the Landau--Silin transport equations.
	In order to illustrate the formalism presented here, we shall give in the accompanying 
article \cite{KupcicJPCM} the detailed description of the CDW mean-field approximation and the effects of 
fluctuations in the order parameter on the dynamical conductivity of the Peierls CDW systems.

\section*{Acknowledgment}
This research was supported by the Croatian Ministry of Science 
and Technology under Project 119-1191458-0512.

\end{document}